\documentclass[conference]{IEEEtran}
\usepackage{textcomp}
\usepackage[utf8]{inputenc}
\usepackage{array}
\usepackage{url}
\usepackage{varwidth}
\usepackage{amsmath}
\usepackage{amssymb}
\usepackage{graphicx}
\usepackage[bookmarks=true,bookmarksnumbered=false,bookmarksopen=false,
 breaklinks=false,pdfborder={0 0 0},pdfborderstyle={},backref=false,colorlinks=false]
 {hyperref}
\hypersetup{pdftitle={BreCol: Benchmarking Classical and Deep-Learning Methods for Microbiome-Based Cancer Detection},
 pdfauthor={Jeffrey M. Dick}}

\makeatletter

\providecommand{\tabularnewline}{\\}

\newcommand{\mytextapprox}{\raisebox{0.5ex}{\texttildelow}}

\usepackage{pbalance}

\makeatother

\begin{document}
\title{BreCol: Benchmarking Classical and Deep-Learning Methods for Microbiome-Based
Cancer Detection}
\author{\IEEEauthorblockN{Jeffrey M. Dick} \IEEEauthorblockA{\textit{School of Geosciences and Info-Physics}\\
 \textit{Central South University}\\
 Changsha, China\\
\href{mailto:jedick@csu.edu.cn}{jedick@csu.edu.cn}}}
\maketitle
\begin{abstract}
DNA sequencing of the gut microbial community shows promise for cancer
detection, but questions remain about the generalizability of results
across studies. We propose BreCol, a benchmark of 2,040 16S rRNA gene
sequencing runs across 26 studies spanning breast cancer, colorectal
cancer, and healthy cohorts. Train-test splits are made within pre-2023
studies, while holdout evaluation uses studies from 2023 onward, reflecting
temporal separation from training data. Classical models reach test/holdout
AUCs of 0.77/0.60 for cancer diagnosis and 1.00/0.83 for cancer type
prediction. We train the models on both cancer types simultaneously
and find that colorectal cancer is often easier to detect than breast
cancer. We also evaluate two deep learning models: HyenaDNA, a long-range
sequence model that pools hidden states for classification, and SetBERT,
a transformer that produces contextualized embeddings over sets of
reads. Both deep learning models underperform the best classical methods
on holdout data, though tuning training set size and the classification
head yields modest gains. Our classical pipeline uses unsupervised
clustering to derive features from tetramer frequencies, preserving
within-run compositional signal and achieving near state-of-the-art
performance without relying on taxonomic assignments. BreCol data
and associated code are publicly available.
\end{abstract}

\begin{IEEEkeywords}
cancer, microbiomes, clustering, benchmarks. 
\end{IEEEkeywords}

\section{Introduction}

The community of microorganisms inhabiting the human digestive tract,
known as the gut microbiome, is increasingly linked to cancer occurrence.
Clinical studies have associated compositional shifts in gut bacteria
with colorectal cancer, and growing evidence implicates gut dysbiosis
in breast cancer as well \cite{YTF+17,ZXS21}. Machine learning models
trained on microbiome profiles have shown promise for distinguishing
cancer patients from healthy controls, raising the prospect of non-invasive,
microbiome-based cancer screening \cite{WPK+19,SHL+25}.

The dominant workflow for characterizing the gut microbiome is 16S
rRNA amplicon sequencing. A short region of the bacterial ribosomal
gene is amplified from fecal samples and sequenced, and the resulting
reads are matched to reference taxa to produce taxon abundance tables.
Most machine learning studies operate on these pre-processed abundance
tables, framing the classification problem as one of finding biomarkers,
i.e. species or genera associated with disease. This discards potentially
informative signal, including fine-grained genetic variation within
taxa, sequences with no close reference in curated databases, and
compositional structure at the level of individual reads within a
sample. Methods that work directly on raw sequence data or on reference-free
sequence features can in principle recover this signal, as shown for
microbiome analysis using \emph{k}-mer frequencies \cite{Bok25}.
Genome language models offer another route to sequence-level representations
and have been benchmarked extensively on genomic tasks \cite{FWZ+25},
but these benchmarks do not target the short, mixed-community reads
produced by 16S rRNA sequencing.

A deeper problem arises when test sets are constructed from the same
studies used for training. This creates optimistically biased performance
estimates that do not reflect real-world deployment, due to both technical
factors (e.g. primer choice and sequencing platform) and regional
microbiome variation \cite{SHL+25,WSNP22}. This problem is exacerbated
for cancer type prediction. Breast and colorectal cancer samples almost
always come from different studies, so a classifier can achieve near-perfect
in-study accuracy simply by identifying the study of origin rather
than the disease. To our knowledge, no previous study has addressed
prediction of cancer type, i.e. distinguishing different primary cancer
sites such as breast and colorectal cancer, from fecal microbiome
data.

\begin{figure*}
\begin{centering}
\includegraphics[width=1\textwidth]{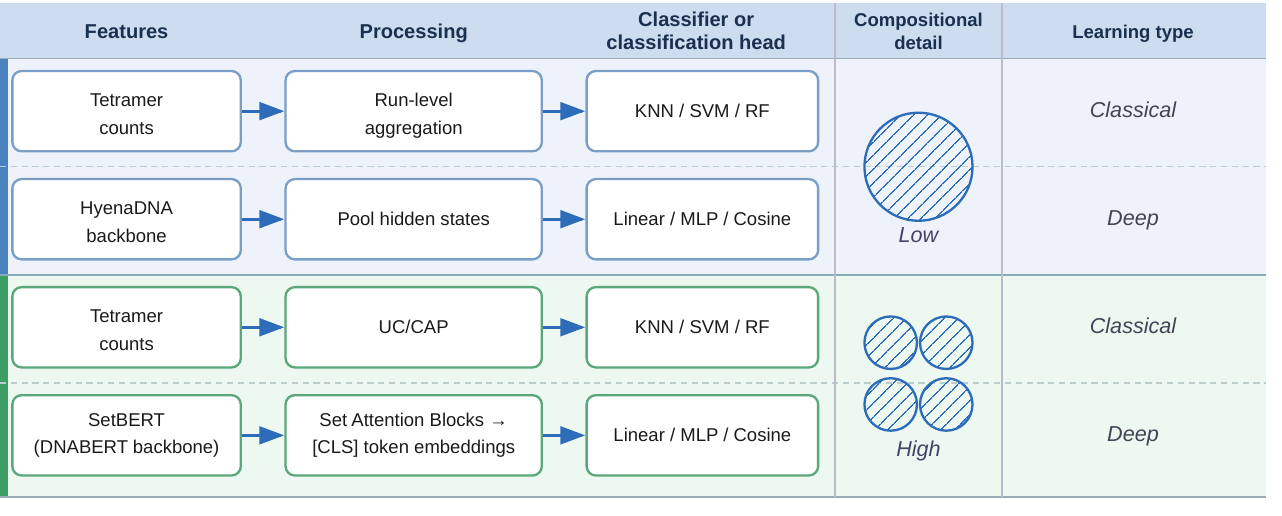}
\par\end{centering}
\caption{Classification pipelines.}\label{fig:pipelines}
\end{figure*}

Reliable benchmarks must evaluate models on one or more external validation
cohorts never encountered during training \cite{WSNP22}; we refer
to these as holdout studies or datasets. Machine learning studies
that draw on published datasets typically reserve only one or two
for external holdout while using the rest for training and internal
cross-validation \cite{TLL+25}. To improve the representation of
holdout datasets, we curated a new compilation of 2,040 16S rRNA sequencing
runs spanning 26 studies (13 breast cancer, 13 colorectal cancer)
from 2013 to 2026. Studies are partitioned chronologically by publication
year; for each cancer type, the first seven studies form the development
partition (training, validation, and test), while the more recent
six studies are reserved as an external holdout.

Against this benchmark we evaluate a progression of feature representations
and learning approaches (Fig.~\ref{fig:pipelines}). For classical
machine learning we use either run-level aggregated tetramer frequencies
or cluster abundance profiles, which serve as an alternative to taxonomic
abundance features without requiring identification of specific taxa
as biomarkers. For deep learning we fine-tune two pre-trained genome
language models: HyenaDNA, a long-context genome foundation model,
and SetBERT, which contextualizes individual reads within their parent
sample. These models were chosen to overcome limitations of other
language models, i.e. short context length, and lack of compositional
representation, which is important for microbial communities. To our
knowledge, this is one of the earliest applications of genome language
models to cancer classification from 16S rRNA microbiome data.

Our main contributions are (1) a custom curated, temporally structured
multi-study benchmark for microbiome-based cancer classification that
provides more reliable estimates of real-world performance than within-study
splits, (2) a reference-free cluster abundance profile method that
achieves the best overall holdout performance among our tested models,
and (3) a comparison of two deep-learning sequence models showing
that both trail the best classical methods on holdout data, with HyenaDNA
generalizing somewhat better than SetBERT.

\section{Methods}

\subsection{Data Curation}

Each sample corresponds to a sequencing run containing multiple 16S
rRNA gene sequences (reads). We searched for gut microbiome studies
for breast cancer or colorectal cancer; studies were only included
if both cancer-positive and healthy control labels were available.
We compiled sample metadata from NCBI BioSample entries and supplementary
information files, then assembled SRA (Sequence Read Archive) run
accessions and downloaded reads from NCBI \cite{SBB+25}.

Our compilation spans 26 studies in total, evenly split between breast
cancer and colorectal cancer (Table~\ref{tab:brecol-data}). Development
studies used for training (including train, validation, and test splits)
and unseen holdout studies are separated not only by study boundaries
but also by time: all holdout studies are from 2023 onward. This design
makes the benchmark a realistic challenge, as predictions must transfer
to future datasets available only after the training date cutoff.

\begin{table*}
\caption{Breast and colorectal cancer studies included in the BreCol compilation.\protect\textsuperscript{}}\label{tab:brecol-data}

\centering{}%
\begin{tabular}{llV{\linewidth}lllllV{\linewidth}l}
\hline 
\textbf{Ref} \textsuperscript{a} & \textbf{Year} & \textbf{BioProject} & \textbf{Type} & \textbf{Cancer} & \textbf{Healthy} & \textbf{Rate} & \textbf{Partition} & \textbf{Country} & \textbf{16S Region}\tabularnewline
\hline 
\cite{AAM+13} & 2013 & PRJNA396901 & breast & 29 & 32 & 1 & development & United States & V4\tabularnewline
\cite{GJH+15} & 2015 & PRJNA345373 & breast & 47 & 47 & 1 & development & United States & V3-V4\tabularnewline
\cite{GHB+18} & 2018 & PRJNA383849 & breast & 48 & 48 & 1 & development & United States & V4\tabularnewline
\cite{BVW+21} & 2021 & PRJNA658160 & breast & 57 & 63 & 0.15 & development & Ghana & V4\tabularnewline
\cite{BSR+22} & 2022 & PRJEB54599 & breast & 19 & 14 & 1 & development & United States & V4\tabularnewline
\cite{WZK+22} & 2022 & PRJNA804967 & breast & 54 & 25 & 1 & development & China & V4\tabularnewline
\cite{ZZZ+22} & 2022 & PRJNA726050 & breast & 14 & 14 & 1 & development & China & V3-V4\tabularnewline
\cite{SKC+23} & 2023 & PRJNA872152 & breast & 22 & 21 & 1 & holdout & United States & V3-V4\tabularnewline
\cite{LBA+25} & 2025 & PRJNA1127492 & breast & 76 & 16 & 1 & holdout & Spain & V2-V3\tabularnewline
\cite{SYL+25} & 2025 & PRJNA1243283 & breast & 10 & 10 & 1 & holdout & China & V3-V4\tabularnewline
\cite{MTK+26} & 2026 & PRJNA914483 & breast & 32 & 32 & 1 & holdout & Malaysia & V3\tabularnewline
\cite{SVK+26} & 2026 & PRJNA1356467 & breast & 22 & 30 & 1 & holdout & India & V3-V4\tabularnewline
\cite{YTK+26} & 2026 & PRJNA1190698 & breast & 15 & 15 & 1 & holdout & Turkey & V1-V9\tabularnewline
\cite{ZTV+14} & 2014 & PRJEB6070 & colorectal & 41 & 75 & 1 & development & France & V4\tabularnewline
\cite{BRRS16} & 2016 & PRJNA290926 & colorectal & 64 & 94 & 0.5 & development & United States and Canada & V4\tabularnewline
\cite{OKN+21} & 2021 & PRJDB11246 & colorectal & 67 & 51 & 0.1 & development & Japan & V1-V2\tabularnewline
\cite{YDS+21} & 2021 & PRJNA763023 & colorectal & 65 & 43 & 0.35 & development & China & V3-V4\tabularnewline
\cite{YWS+21} & 2021 & PRJEB36789 & colorectal & 53 & 52 & 1 & development & Argentina, Chile,

India, and Vietnam & V4\tabularnewline
\cite{DLT+22} & 2022 & PRJNA824020 & colorectal & 27 & 33 & 1 & development & China & V4\tabularnewline
\cite{PCL+22} & 2022 & PRJNA662014 & colorectal & 36 & 25 & 1 & development & Singapore & V3-V4\tabularnewline
\cite{BWY+23} & 2023 & PRJEB53415 & colorectal & 46 & 43 & 1 & holdout & India & V4\tabularnewline
\cite{BRR+24} & 2024 & PRJEB71787 & colorectal & 51 & 51 & 1 & holdout & Spain & V3-V4\tabularnewline
\cite{CAB+24} & 2024 & PRJNA911189 & colorectal & 90 & 30 & 1 & holdout & Spain & V3-V4\tabularnewline
\cite{SGH+24} & 2024 & PRJNA1059759 & colorectal & 10 & 10 & 1 & holdout & India & V3-V4\tabularnewline
\cite{ARF+25} & 2025 & PRJEB76625 & colorectal & 25 & 15 & 1 & holdout & Iran & V3-V4\tabularnewline
\cite{GYX+25} & 2025 & PRJNA1092526 and

PRJNA1092376 & colorectal & 67 & 64 & 0.6 & holdout & China & V3-V4\tabularnewline
\hline 
\multicolumn{10}{>{\raggedright}p{0.9\textwidth}}{\textsuperscript{a}Studies are arranged chronologically by publication
year and divided into development and holdout partitions. Cancer and
healthy numbers reflect counts after stratified subsampling at the
indicated rate.}\tabularnewline
\end{tabular}
\end{table*}

Some studies have substantially larger sample counts than others.
To improve study balance, we applied random subsampling within several
studies (stratified by cancer vs healthy label). The sample sizes
in Table~\ref{tab:brecol-data} reflect counts after sampling at
the indicated rate (fraction of samples). Additionally, for two studies
we excluded runs with \textless 2000 reads \cite{BVW+21,CAB+24}.

\subsection{Preprocessing, Splits, and Sampling}

We normalized sample labels using three classes: healthy, breast cancer,
and colorectal cancer. Breast cancer samples include invasive tumors;
colorectal cancer samples include carcinoma. Any benign samples (e.g.
adenomas, benign colon polyps, and breast ductal carcinoma in situ
(DCIS)) and non-fecal samples in the datasets were excluded from our
analysis. Among development studies, we assigned each sequencing run
to stratified training, validation, or test sets in a 70:10:20 ratio.
Split assignments were defined in advance, independent of any downstream
feature computation. Runs from holdout studies were excluded from
this assignment. 

We used a single fixed validation split rather than \emph{k}-fold
cross-validation, since repeated GPU-intensive model training would
be prohibitively expensive. This allows the same development splits
to be used consistently across both the classical and deep-learning
pipelines. The same run-level split underlies both classification
tasks: cancer versus healthy (cancer diagnosis) on all samples, and
breast versus colorectal (cancer type) restricted to cancer-positive
samples. 

For all classification pipelines we dropped the first 1000 sequences
in each run to avoid potential quality artifacts from run initiation.
For classical feature computation, we then randomly sampled up to
5000 sequences from the remaining sequences in each run. For deep-learning
models, smaller sequence sets were used, as described below.

\subsection{Classical Machine Learning}

\subsubsection{Run-level Tetramer Frequencies}

Tetramer frequencies, also known as tetranucleotide frequencies, have
proven useful for clustering metagenomic sequences \cite{DAB+09}
and for microbiome-level genomic differentiation \cite{KBC+25}. We
calculated tetramer frequencies for each run by counting all 4-mers
within each sequence sampled from the run, summing counts over all
sampled sequences, then converting to relative frequencies, yielding
a 256-dimensional feature vector used for supervised classification.

\subsubsection{Cluster Abundance Profiles for Tetramer Counts}

Because run-level tetramer frequencies are averages, they lose within-run
compositional structure, i.e. information about which sequence types
tend to co-occur in the same sample. Taxonomic profiling preserves
this structure through genus- or species-level groupings, but taxonomic
assignment is not the only route to sequence similarity clusters.

To preserve within-run compositional structure, we used unsupervised
clustering followed by cluster abundance profiles (UC/CAP). Even with
mini-batch \emph{k}-means, clustering all available sequences across
training runs exhausts the RAM on our machine. Therefore, we performed
unsupervised clustering by drawing at most a fixed number of sequences
per training run (\emph{n}\textsubscript{UC}). For each selected
sequence we computed the 256-dimensional tetramer composition vector,
then fit mini-batch \emph{k}-means to all selected sequences to obtain
\emph{K} centroids defining a sequence codebook. Dimensionality reduction
with PCA before \emph{k}-means was trialed and found to degrade downstream
classification results, so it was not used here.

To construct run-level features, we applied the same centroid assignments
(without refitting) to a larger per-run sequence budget (\emph{n}\textsubscript{CAP}),
including validation, test, and holdout runs. We counted cluster memberships
within each run and normalized by the number of assigned sequences
to produce a \emph{K}-dimensional cluster abundance profile (CAP).
These CAP vectors serve as the feature matrix for supervised classification.

\subsubsection{Classification Pipeline}

For the majority-class baseline, we predicted the most frequent class
in the training set. For other classifier models we performed grid
search over the hyperparameters listed in Table~\ref{tab:hyperparameters}.

For both KNN and SVM we applied a centered log-ratio transform (CLR)
to handle compositional data, standardized the CLR coordinates, then
applied principal component analysis (PCA) to reduce dimensionality.
For KNN we used inverse distance weighting and tuned the PCA components
and number of neighbors. For SVM, we used an RBF kernel and tuned
the PCA components and penalty parameter \emph{C}. The kernel width
parameter \emph{gamma} was left at scikit-learn's default (\texttt{scale}).
For random forest, we used the same CLR and standardization but omitted
PCA. We tuned the number of trees, maximum tree depth, and minimum
samples per leaf.

After selecting hyperparameters using area under the receiver operating
characteristic (ROC) curve (AUC) by grid search on the validation
split, we fit each final pipeline on the training split.

\begin{table}
\caption{Classifier models and hyperparameter grids used with run-level tetramer
frequencies and cluster abundance profiles (UC/CAP features).}\label{tab:hyperparameters}

\centering{}%
\begin{tabular}{lV{\linewidth}}
\hline 
\textbf{Model} & \textbf{Hyperparameters}\tabularnewline
\hline 
KNN & PCA n\_components (none, 0.95),

n\_neighbors (5, 15)\tabularnewline
SVM & PCA n\_components (none, 0.95),

C (1.0, 10.0)\tabularnewline
Random Forest & n\_estimators (200, 500),

max\_depth (none, 10),

min\_samples\_leaf (1, 2)\tabularnewline
\hline 
\end{tabular}
\end{table}

\subsection{Deep Learning Models}

\textbf{HyenaDNA} \cite{NPF+23} is a long-range genomic foundation
model pre-trained on the human reference genome. It uses single-nucleotide
tokens and a stack of Hyena operators for implicit convolutions to
process sequences of up to 1 million bases. For each sequencing run
we pack sequences into sets and mean-pool the per-position hidden
states of the backbone to produce a single vector for classification.
This follows the pooling strategy used by the HyenaDNA authors for
their 1k and 32k models on a species classification task \cite{NPF+23}.

\textbf{SetBERT} \cite{LGA+25} is a transformer architecture designed
for high-throughput sequencing data. It represents each read with
a DNABERT sequence encoder using overlapping 3-mer (trinucleotide)
tokens and then contextualizes the sequences from a given sample with
a stack of Set Attention Blocks (SABs). SABs are standard transformer
blocks without positional encoding, making the model permutation-equivariant.
A learned class token ({[}CLS{]}) prepended to the set is conditioned
on all reads by the SABs; its output embedding summarizes the entire
run and serves as input to the classification head. SetBERT was pre-trained
on approximately 280,000 microbial 16S rRNA amplicon samples with
a relative-abundance prediction objective.

For both models we tested three classification head architectures
applied to the run-level summary vector. A \textbf{linear} head is
a single fully connected layer mapping the embedding to a scalar logit.
This is the simplest option and is the decoder head used in the original
HyenaDNA paper; it is also the closest to the pre-trained regime for
SetBERT (linear head with softmax for relative abundance prediction).
An \textbf{MLP} (multi-layer perceptron) head adds a hidden layer
(in our study: 256 units, GELU activation, 0.1 dropout) before the
output layer, giving the classifier more capacity to learn non-linear
decision boundaries. A \textbf{cosine similarity} head projects the
embedding onto a single learned direction, scoring it by cosine similarity
scaled by a learnable temperature. Because cosine similarity is direction-only,
it is sensitive to the orientation of the embedding vector rather
than its magnitude.

\subsection{HyenaDNA and SetBERT Fine-Tuning}

We initialized HyenaDNA from the \emph{hyenadna-small-32k-seqlen}
pre-trained checkpoint and fine-tuned separate models for each task.
We use binary cross-entropy loss averaged across all sequence sets
for each run. At evaluation, set-level logits were averaged to obtain
one prediction per run. 

For SetBERT, we used the released \emph{qiita-16s} checkpoint: a 12-layer
DNABERT encoder embeds each amplicon read into a 768-dimensional vector,
and a 6-layer SAB transformer with 12 heads contextualizes the set
of read embeddings, producing a {[}CLS{]} token embedding that summarizes
the run. For each task we attached a classification head to the {[}CLS{]}
token and trained with binary cross-entropy.

Although HyenaDNA offers large context lengths (up to 1 million bases)
and SetBERT was pre-trained with 1000 sequences per run and tested
with up to 10,000 sequences per run \cite{LGA+25}, the 16 GB of GPU
RAM available for our study required using relatively few sequences.
For HyenaDNA we packed non-truncated sequences from each run into
the context window (up to 16k positions in our experiments) and generated
5 sets of packed sequences per run, representing up to \mytextapprox323
sequences per run. For SetBERT, we trimmed sequences to 150 bp (following
the SetBERT paper) then used 350 sequences per run. 

For both models we used a learning rate of 1\texttimes 10\textsuperscript{\textminus 4}
for the classification head and 1\texttimes 10\textsuperscript{\textminus 5}
for the backbone (a 10\texttimes{} reduction to preserve pre-trained
representations). The optimizer was AdamW and weight decay was set
to 0.1. Models were trained for 5 epochs, and the epoch with the highest
validation AUC was selected. 

\subsection{Implementation}

The benchmark dataset is composed of CSV files including accession
numbers and metadata, together with scripts for downloading sequence
data from NCBI. The project code is written in Python with YAML configuration
files and a Makefile-driven analysis pipeline. The official HyenaDNA
and SetBERT implementations were modified for this project and structured
as pip-installable packages for import by the analysis scripts. After
downloading the data, the entire pipeline runs in approximately 25
hours on a machine with 8 CPU cores, a 16 GB NVIDIA GPU, and 32 GB
of RAM.

\section{Results}

We define two binary classification tasks: \textbf{cancer diagnosis}
(cancer vs. healthy, all samples for both cancer types) and \textbf{cancer
type} (breast vs. colorectal, cancer-positive samples only). Performance
is reported as AUC on the test split (unseen samples from the development
studies used to train the model) and the holdout split (entirely unseen
studies).

\subsection{Classification with Run-Level Tetramer Frequencies}

All models exceed the majority-class baseline on the test split, with
particularly large margins for cancer type prediction (Table~\ref{tab:tetramer}).
The holdout picture is sharply different. For cancer diagnosis, SVM
achieves the highest AUC of 0.59 followed closely by KNN and random
forest. For cancer type, SVM reaches an AUC of 0.71 on holdout followed
by random forest, while KNN collapses below baseline.

\subsection{Classification with cluster abundance profiles for tetramer counts}

We explored six combinations of the three UC/CAP hyperparameters defined
by \emph{n}\textsubscript{UC} (sequences per run used for unsupervised
clustering), \emph{K} (number of clusters), and \emph{n}\textsubscript{CAP}
(sequences per run assigned to centroids and used to build cluster
abundance profiles) (Table~\ref{tab:feature_sets}).

These UC/CAP parameters produced six different cluster abundance profiles
(or feature sets) used for standard supervised classification. SVM
achieves higher holdout AUC than KNN across feature sets for cancer
diagnosis, but the pattern is reversed for cancer type, where KNN
leads (Fig.~\ref{fig:tetramer_uc_cap}). For cancer type, both models
show near-perfect in-study test performance across feature sets, but
holdout values drop sharply.

Table~\ref{tab:tetramer_uc_cap} lists the results for the best UC/CAP
feature set as judged by test AUC across models in each task. Selecting
on the test AUC of development studies does not invalidate the holdout
AUC, which represents performance on entirely unseen datasets. For
cancer diagnosis, SVM achieves the best holdout performance, followed
by random forest and KNN. For cancer type, KNN leads on holdout, followed
by random forest and SVM. The gap between in-study test and holdout
is again large for cancer type, but UC/CAP with KNN achieves substantially
higher cancer type holdout AUC than any classifier based on run-level
tetramer frequencies.

\subsection{Classification with HyenaDNA}

For each task we fine-tuned a HyenaDNA model from the pre-trained
backbone with different head architectures. For cancer diagnosis,
test and holdout AUC are similar across all three heads, at around
0.61--0.62 for test and 0.54--0.57 for holdout (Table~\ref{tab:hyenadna}).
For cancer type, all heads show similar test AUC (0.88--0.89) and
the MLP head achieves the highest holdout AUC by a small margin (0.79),
though with the largest standard deviation.

\begin{table}
\caption{Test and holdout AUC for run-level tetramer frequencies. Bold marks
the best value per column.}\label{tab:tetramer}

\centering{}%
\begin{tabular}{ccccc}
 & \multicolumn{2}{c}{\textbf{Cancer diagnosis}} & \multicolumn{2}{c}{\textbf{Cancer type}}\tabularnewline
\hline 
\textbf{Model} & \textbf{\emph{Test}} & \textbf{\emph{Holdout}} & \textbf{\emph{Test}} & \textbf{\emph{Holdout}}\tabularnewline
\hline 
Majority class & 0.50 & 0.50 & 0.50 & 0.50\tabularnewline
KNN & 0.71 & 0.58 & 0.96 & 0.41\tabularnewline
SVM & 0.70 & \textbf{0.59} & \textbf{0.996} & \textbf{0.71}\tabularnewline
Random forest & \textbf{0.73} & 0.57 & 0.99 & 0.66\tabularnewline
\hline 
\end{tabular}
\end{table}

\begin{table}
\caption{UC/CAP feature sets.}\label{tab:feature_sets}

\centering{}%
\begin{tabular}{cccccccc}
\hline 
Set & \emph{n}\textsubscript{UC} & \emph{K} & \emph{n}\textsubscript{CAP} & Set & \emph{n}\textsubscript{UC} & \emph{K} & \emph{n}\textsubscript{CAP}\tabularnewline
\hline 
1 & 350 & 1000 & 350 & 4 & 1000 & 1000 & 5000\tabularnewline
2 & 1000 & 1000 & 1000 & 5 & 1000 & 2000 & 5000\tabularnewline
3 & 1000 & 2000 & 1000 & 6 & 1000 & 3000 & 5000\tabularnewline
\hline 
\end{tabular}
\end{table}

\begin{figure}
\begin{centering}
\includegraphics[width=1\columnwidth]{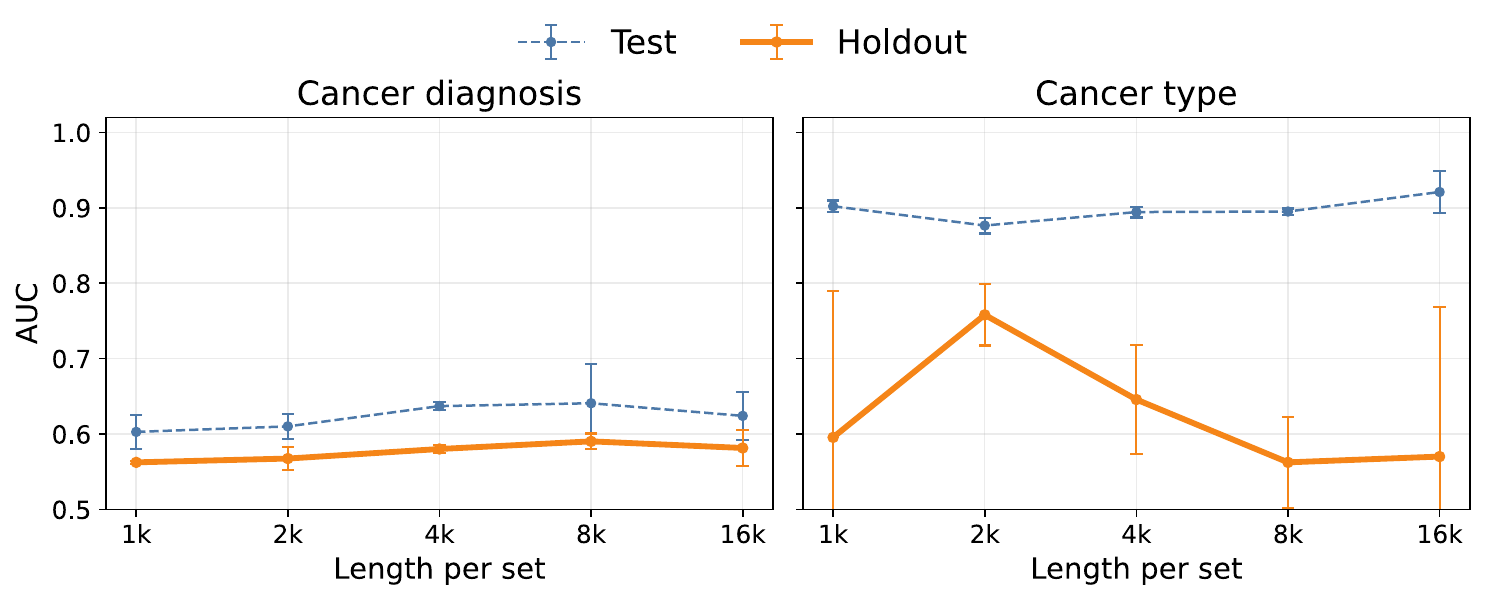}
\par\end{centering}
\caption{Test and holdout AUC for HyenaDNA with the linear classification
head and varying sequence set length. Error bars represent standard
deviation with three random seeds.}\label{fig:hyenadna}
\end{figure}

We used the linear head (the configuration with the best holdout AUC
for cancer diagnosis) to study the effect of context length. We varied
the length per sequence set (1k, 2k, 4k, 8k, and 16k positions), obtaining
shorter configurations by truncating a single large cache built at
16k. Fig.~\ref{fig:hyenadna} shows that for cancer diagnosis, holdout
AUC increases modestly from 2k to 8k before leveling off, while for
cancer type, contexts longer than 2k positions markedly reduce holdout
AUC despite slight gains on the test split. The divergence between
test and holdout trends suggests that larger contexts allow the model
to pick up study-specific signals for cancer type prediction.

\begin{table}
\caption{Test and holdout AUC for cluster abundance profiles. The best UC/CAP
feature set per task was selected by test AUC.}\label{tab:tetramer_uc_cap}

\centering{}%
\begin{tabular}{ccccc}
 & \multicolumn{2}{c}{\textbf{Cancer diagnosis}} & \multicolumn{2}{c}{\textbf{Cancer type}}\tabularnewline
\hline 
\textbf{Model} & \textbf{\emph{Test}} & \textbf{\emph{Holdout}} & \textbf{\emph{Test}} & \textbf{\emph{Holdout}}\tabularnewline
\hline 
 & \multicolumn{2}{c}{Feature set 5} & \multicolumn{2}{c}{Feature set 5}\tabularnewline
KNN & 0.69 & 0.54 & 0.998 & \textbf{0.83}\tabularnewline
SVM & \textbf{0.77} & \textbf{0.60} & \textbf{1.00} & 0.74\tabularnewline
Random forest & 0.74 & 0.57 & 1.00 & 0.79\tabularnewline
\hline 
\end{tabular}
\end{table}

\begin{figure}
\begin{centering}
\includegraphics[width=1\columnwidth]{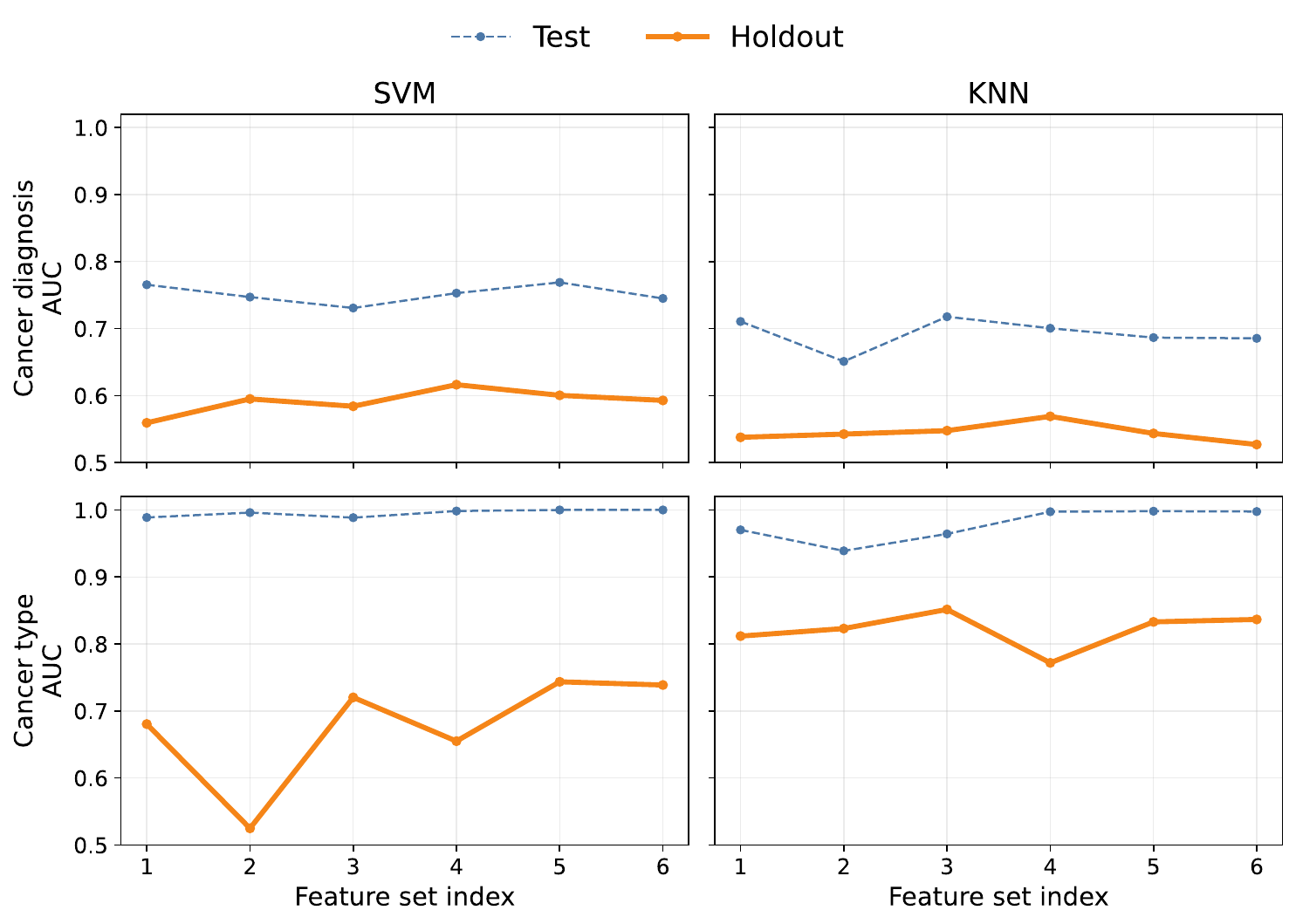}
\par\end{centering}
\caption{Test and holdout AUC for SVM and KNN models trained with cluster
abundance profiles. UC/CAP feature sets are listed in Table~\ref{tab:feature_sets}.}\label{fig:tetramer_uc_cap}
\end{figure}

For a fair comparison between HyenaDNA and the UC/CAP pipeline, we
select configurations with similar sequence counts per sample. UC/CAP
feature set 1 uses 350 sequences per sample for clustering (Table~\ref{tab:feature_sets}).
For HyenaDNA at 16k positions per set and 5 sets per run, the number
of sequences per sample is 323 \textpm{} 112 (min 50 for \cite{YTK+26},
max 540 for \cite{BVW+21}). The results for these settings are visualized
for UC/CAP in Fig.~\ref{fig:tetramer_uc_cap} (feature set 1) and
for HyenaDNA in Fig.~\ref{fig:hyenadna} (16k set length). For cancer
diagnosis, HyenaDNA loses to both SVM and KNN on test AUC, but shows
competitive holdout AUC near 0.58. For cancer type, HyenaDNA shows
respectable test AUC (\textgreater 0.9) but struggles on holdout
(\textless 0.6), considerably lower than either SVM or KNN.

\subsection{Classification with SetBERT}

We evaluated the same three classification heads on SetBERT (Table~\ref{tab:setbert}).
For cancer diagnosis, test and holdout AUC are similar across heads
(0.61--0.65 test, 0.55--0.58 holdout), as with HyenaDNA. For cancer
type, test AUC is again similar across heads (0.97--0.98), while
the cosine similarity head is best on holdout (0.71) and the MLP head
is the worst on holdout (0.62), with substantially higher variance
than the other two heads.

\begin{table}
\caption{HyenaDNA fine-tuning results with 5 sets at 2048 max length per set,
reported as mean \textpm{} standard deviation across three random
seeds.}\label{tab:hyenadna}

\centering{}%
\begin{tabular}{ccccc}
 & \multicolumn{2}{c}{\textbf{Cancer diagnosis}} & \multicolumn{2}{c}{\textbf{Cancer type}}\tabularnewline
\hline 
\textbf{Head} & \textbf{\emph{Test}} & \textbf{\emph{Holdout}} & \textbf{\emph{Test}} & \textbf{\emph{Holdout}}\tabularnewline
\hline 
Linear & 0.61 \textpm{} 0.02 & \textbf{0.57} \textpm{} 0.02 & 0.88 \textpm{} 0.01 & 0.74 \textpm{} 0.07\tabularnewline
MLP & 0.61 \textpm{} 0.04 & 0.54 \textpm{} 0.03 & 0.88 \textpm{} 0.02 & \textbf{0.79} \textpm{} 0.10\tabularnewline
Cosine & \textbf{0.62} \textpm{} 0.01 & 0.55 \textpm{} 0.02 & \textbf{0.89} \textpm{} 0.01 & 0.74 \textpm{} 0.08\tabularnewline
\hline 
\end{tabular}
\end{table}

\begin{table}
\caption{SetBERT fine-tuning results with a per-run set size of 350 sequences,
reported as mean \textpm{} standard deviation across three random
seeds.}\label{tab:setbert}

\centering{}%
\begin{tabular}{ccccc}
 & \multicolumn{2}{c}{\textbf{Cancer diagnosis}} & \multicolumn{2}{c}{\textbf{Cancer type}}\tabularnewline
\hline 
\textbf{Head} & \textbf{\emph{Test}} & \textbf{\emph{Holdout}} & \textbf{\emph{Test}} & \textbf{\emph{Holdout}}\tabularnewline
\hline 
Linear & 0.61 \textpm{} 0.07 & 0.55 \textpm{} 0.02 & 0.98 \textpm{} 0.01 & 0.66 \textpm{} 0.04\tabularnewline
MLP & 0.61 \textpm{} 0.02 & \textbf{0.58} \textpm{} 0.04 & \textbf{0.98} \textpm{} 0.01 & 0.62 \textpm{} 0.15\tabularnewline
Cosine & \textbf{0.65} \textpm{} 0.02 & 0.56 \textpm{} 0.02 & 0.97 \textpm{} 0.00 & \textbf{0.71} \textpm{} 0.04\tabularnewline
\hline 
\end{tabular}
\end{table}

\section{Discussion}

Results are consistently lower on holdout splits than on in-study
test splits. For run-level tetramer frequencies, the stark contrast
between test and holdout performance (AUC \textgreater 0.9 for test
vs 0.71 or less for holdout) indicates that classifiers overfit to
study-level signals for the cancer type task. Comparing holdout performance
across Tables~\ref{tab:tetramer} and \ref{tab:tetramer_uc_cap},
UC/CAP offers a consistent advantage over run-level tetramer features
for cancer type. However, it barely improves, and in some cases even
weakens, holdout AUC for cancer diagnosis on in-study test splits
and external holdout studies. This suggests that the within-run compositional
structure captured by cluster abundance profiles partially breaks
the study-level shortcuts that hinder cancer type classifiers, but
may not generate transferable signal for cancer diagnosis.

We list per-study AUCs for cancer diagnosis using UC/CAP and comparisons
with published values for colorectal cancer where available (Table~\ref{tab:auc_comparison}).
On two of the three development studies with published values \cite{ZTV+14,YDS+21},
our test AUC is relatively high (0.98--1.00), but it drops to 0.73
on a third dataset where the literature value is 0.85 \cite{BRRS16}.
For holdout studies with published AUC values \cite{BWY+23,CAB+24,GYX+25},
our AUC (0.66--0.68) is consistently lower than the literature (0.86--0.88).
The literature numbers come from within-study cross-validation or
test splits rather than independent cohorts, explaining their higher
values compared to holdout performance.

We did not find microbiome-based AUC values reported in the breast
cancer studies we used, but some comparable results exist. Wang et
al. \cite{WYH+22} trained random forest classifiers on amplicon sequence
variants using 16S rRNA data from breast tissue and fecal microbiomes;
two of the latter correspond to development datasets in our compilation
\cite{GHB+18,BVW+21}. While Wang et al. reported higher cancer-diagnosis
AUCs for tissue than for fecal microbiomes, their cross-cohort AUCs
of 0.54--0.59 for the two fecal datasets are in the range of our
holdout values for breast cancer diagnosis (0.47--0.69). 

\begin{table}
\caption{Per-study AUC for cancer diagnosis in this study compared with literature
values.\protect\textsuperscript{}}\label{tab:auc_comparison}

\centering{}%
\begin{tabular}{cccccccc}
 & \multicolumn{3}{c}{\textbf{Breast cancer}} & \multicolumn{4}{c}{\textbf{Colorectal cancer}}\tabularnewline
\hline 
\textbf{Partition} & \textbf{\emph{Ref}} & \textbf{\emph{n}} & \textbf{\emph{AUC }}\textsuperscript{a} & \textbf{\emph{Ref}} & \textbf{\emph{n}} & \textbf{\emph{AUC}} & \textbf{\emph{Lit}}\tabularnewline
\hline 
Development & \cite{AAM+13} & 9 & 0.21 & \cite{ZTV+14} & 23 & 0.98 & 0.84\tabularnewline
Development & \cite{GJH+15} & 19 & 0.72 & \cite{BRRS16} & 31 & 0.73 & 0.85\tabularnewline
Development & \cite{GHB+18} & 20 & 0.62 & \cite{OKN+21} & 23 & 0.35 & ---\tabularnewline
Development & \cite{BVW+21} & 29 & 0.71 & \cite{YDS+21} & 18 & 1.00 & 0.87\tabularnewline
Development & \cite{BSR+22} & 7 & 0.60 & \cite{YWS+21} & 20 & 0.92 & ---\tabularnewline
Development & \cite{WZK+22} & 15 & 0.52 & \cite{DLT+22} & 16 & 0.78 & ---\tabularnewline
Development & \cite{ZZZ+22} & 8 & 1.00 & \cite{PCL+22} & 10 & 0.88 & ---\tabularnewline
Holdout & \cite{SKC+23} & 43 & 0.58 & \cite{BWY+23} & 89 & 0.66 & 0.86\tabularnewline
Holdout & \cite{LBA+25} & 92 & 0.51 & \cite{BRR+24} & 102 & 0.61 & ---\tabularnewline
Holdout & \cite{SYL+25} & 20 & 0.50 & \cite{CAB+24} & 120 & 0.68 & 0.86\tabularnewline
Holdout & \cite{MTK+26} & 64 & 0.47 & \cite{SGH+24} & 20 & 0.52 & ---\tabularnewline
Holdout & \cite{SVK+26} & 52 & 0.68 & \cite{ARF+25} & 40 & 0.70 & ---\tabularnewline
Holdout & \cite{YTK+26} & 30 & 0.69 & \cite{GYX+25} & 131 & 0.68 & 0.88\tabularnewline
\hline 
\multicolumn{8}{>{\raggedright}p{0.95\columnwidth}}{\textsuperscript{a}AUC from the best tetramer UC/CAP classifier in
Table~\ref{tab:tetramer_uc_cap} (SVM, feature set 5). AUC is computed
over each study's test-split runs (development) or all runs (holdout);
the \emph{n} column reports the total number of samples (cancer +
healthy) contributing to each per-study AUC. Literature AUC values
for colorectal cancer are shown where reported.}\tabularnewline
\end{tabular}
\end{table}

Interestingly, the test AUCs for breast cancer are on average lower
than those for colorectal cancer (means of 0.63 vs 0.81 for the development
studies in Table~\ref{tab:auc_comparison}). This pattern extends
to the holdout studies; for breast cancer only two out of six holdout
studies have AUC \textgreater{} 0.6, while for colorectal cancer this
grows to five out of six studies. This indicates easier detection
of colorectal cancer than breast cancer when models are simultaneously
trained on data from both cancer types, as done in this study.

\subsection{HyenaDNA versus SetBERT}

Both deep-learning models underperform the best classical methods
on holdout data. Comparing Tables~\ref{tab:hyenadna} and \ref{tab:setbert},
the two models produce nearly identical holdout AUC for cancer diagnosis
(best: 0.57 for HyenaDNA linear, 0.58 for SetBERT MLP). For cancer
type, SetBERT achieves higher test AUC (0.98 versus 0.89 for HyenaDNA)
but HyenaDNA generalizes better to holdout studies (best: 0.79 for
HyenaDNA with MLP versus 0.71 for SetBERT with cosine similarity head).
HyenaDNA's stronger holdout AUC on cancer type is notable given that
it was pre-trained on the human genome rather than on microbial sequences;
the domain mismatch does not appear to be the limiting factor.

The aggregated representation in HyenaDNA before classification discards
within-run compositional structure, which UC/CAP explicitly preserves.
SetBERT is an interesting alternative as it uses set attention blocks
so embeddings are affected by sample context (i.e. other sequences).
In our experiments, SetBERT performs better than HyenaDNA on in-study
test splits but not on holdout datasets. A possible contributing factor
is that SetBERT sequence processing takes only 150 bp from each sequence,
which may lose information useful for generalizing to unseen studies.
Also, SetBERT was pre-trained on V3-V4 regions of 16S rRNA, while
some of our holdout studies use different regions.

\subsection{Limitations and directions for improvement}

Several limitations should be noted. First, the cancer-type task combines
female-only datasets (breast cancer) with datasets of mixed sex (colorectal
cancer). Sex-specific differences in fecal microbiome composition
\cite{GVR25} could confound this comparison. Sex metadata (male/female)
are available for at least six of the 13 colorectal cancer studies
in our compilation \cite{ZTV+14,BRRS16,YWS+21,DLT+22,CAB+24,ARF+25},
enabling future filtering to female-only participants for controlled
cancer-type comparisons. Second, both deep-learning models were fine-tuned
with a small fraction of available sequences per run. Strategies that
use more sequences could better exploit available data. However, our
experiments do not support set size between 1k and 16k as the limiting
factor for HyenaDNA (Fig.~\ref{fig:hyenadna}). Third, our reference-free
tetramer and cluster abundance features, as well as the deep-learning
embeddings, are not interpretable as taxonomic biomarkers, limiting
biological interpretation of which microbial taxa drive classification
decisions. Finally, no baseline values for classification using traditional
taxonomic abundance features were obtained in this study. Comparisons
with previous work (\cite{ZTV+14,YDS+21} in Table~\ref{tab:auc_comparison})
suggest that UC/CAP has the ability to surpass taxonomy-based classifiers
on in-study test splits, but a systematic baseline on all datasets
in the benchmark would be more conclusive.

Several avenues may improve holdout performance. UC/CAP parameters
(\emph{n}\textsubscript{UC},\emph{ K}, \emph{n}\textsubscript{CAP})
could be tuned jointly with the classifier rather than selected independently.
Soft cluster assignments (Gaussian mixture or fuzzy \emph{k}-means)
might better capture the continuous composition of microbial communities.
Table~\ref{tab:auc_comparison} reveals substantial variation in
per-study AUC for cancer diagnosis; targeting the most challenging
studies for model improvement, for example, by up-weighting hard examples
during training, could be a productive direction for future work.
For HyenaDNA, pre-training on microbial 16S rRNA sequences would better
align its representations with the target domain.

\section*{Acknowledgments}

\noindent No funding was received for conducting this study. The author
has no relevant financial or nonfinancial interests to disclose.

\section*{Declaration of Generative AI Use}

\noindent Cursor was used for code generation. Cursor and Claude Sonnet
4.6 were used for writing sections of the manuscript. Claude Sonnet
4.6 and 5 were used for polishing the text. AI-generated text was
incorporated into the manuscript after human review and cleanup by
the author. Manuscript revision history and prompts are available
in the project repository.

\section*{Data and Code Availability}

\noindent The data compilation and code produced in this study are
available at \url{https://github.com/jedick/BreCol}.

\bibliographystyle{IEEEtranDOI}
\bibliography{references}

\end{document}